\documentclass[useAMS,usenatbib]{mn2e}
\usepackage{graphicx}

\title[Composite Interstellar Grains]
{Composite Interstellar Grains}
\author[Vaidya et al.]{D.B. Vaidya$^{1}$\thanks{E-mail:
dbv@satyam.net.in; rag@iucaa.ernet.in and Theodore.Snow@Colorado.EDU}, Ranjan Gupta$^{2}$ and
T.P. Snow$^{3}$
\\
$^{1}$Gujarat College, Ahmedabad-380006, India\\
$^{2}$IUCAA, Post Bag 4, Ganeshkhind, Pune-411007, India\\
$^{3}$Center for Astrophysics \& Space Astronomy, University
 of Colorado, Boulder, CO-80309-0389, USA\\}
\begin{document}

\date{Received on /10/2006}

\pagerange{\pageref{firstpage}--\pageref{lastpage}} \pubyear{2006}

\maketitle

\label{firstpage}

\begin{abstract}

A composite dust grain model which is consistent with
the observed interstellar extinction and linear polarization
is presented.
The composite grain is made up
of a host silicate spheroid and graphite inclusions.
The extinction
efficiencies of the composite spheroidal grains for three axial
ratios are computed using the discrete dipole approximation (DDA).
The interstellar extinction curve is evaluated in the spectral
region  3.40--0.10$\mu m$ using the extinction efficiencies of the
composite spheroidal grains. 
The model extinction curves are then compared with the average
observed interstellar extinction curve. We also calculate the linear
polarization for the spheroidal composite grains at three
orientation angles and find the wavelength of maximum
polarization. 
Further, we estimate the volume extinction factor, an
important parameter from the point of view of cosmic abundance,
for the composite grain models that reproduce the average observed
interstellar extinction. The estimated abundances derived from the composite
grain models for both carbon
and silicon are found to be lower than that are predicted by the bare
silicate/graphite grain models but these values are still higher than
that are implied from the recent ISM values.
\end{abstract}

\begin{keywords}
Interstellar Dust, Extinction, Linear Polarization, 
Cosmic Abundances
\end{keywords}

\section{Introduction}
The most commonly used interstellar dust grain model, consists of
two distinct populations of bare spherical silicate and graphite grains
with a power law size distribution (Mathis et al. 1977) .
This model, popularly known as MRN model provides an excellent fit
to the average observed interstellar extinction curve.
However, it is highly unlikely that the interstellar grains are
spherical in shape or that they are homogeneous in composition and
structure. The collected interplanetary particles are nonspherical and highly
porous and composites of very small sub-grains glued together (Brownlee,
1987). Moreover, the interstellar polarization that accompanies extinction
requires that the interstellar grains must be aligned and nonspherical (Wolff et.
al. 1993).
Recently Stark et al. (2006) have suggested the growth of spheroidal grains via
plasma deposition in the supernova remnant.
The elemental abundances derived from the observed
interstellar extinction also do not favour the homogeneous
composition for the interstellar grains. Recent studies suggest that the
abundances of various elements in the interstellar medium (ISM) are
less than their solar values (Snow \& Witt, 1996; Voshchinnikov 
et al., 2006).
For example in case of carbon, its solar abundance
normalized to a hydrogen abundance of $10^{6}$ atoms, is about 355
(Grevesse et al. 1996). In the ISM the abundance of this element is
significantly lower, about 225 C atoms (Snow \& Witt 1995).
Also a significant amount of this interstellar carbon is in gas phase.
From CII 2325\AA~ absorption measurements,
Cardelli et al. (1996) have inferred a C abundance of
about 150 atoms . This means that a total of about 100 C atoms are
available for the dust phase, compared with about 300 atoms
required for the bare grain models (e.g. Mathis et al. 1977, MRN).
In order to overcome the cosmic abundance constraints Mathis \&
Whiffen (1989), Mathis (1996) and Dwek (1997) have proposed composite
 grain models consisting of silicate and amorphous carbon as
constituent materials. They have used effective medium
theory (EMT) to calculate the optical constants for composite
grains and then used the Mie theory to calculate extinction cross sections
for spheres In EMT the inhomogeneous particle is replaced by a
homogeneous one with some 'average effective dielectric function'.
The effects related to the fluctuations of the dielectric function within
the inhomogeneous structures cannot be treated by this approach
of the EMT. Mathis (1996) has also noted the uncertainty in the use of EMT for
treating the composite grains and has suggested that detailed calculations such
as the DDA would be necessary for the treatment of voids in silicate/graphite
particles for some wavelengths.

Iati et al. (2004) have studied optical properties of composite grains
as grain aggregates of amorphous carbon and astronomical silicates, 
using the transition matrix approach. Voshchinnikov et al. (2005)
have studied properties of composite grains as layered spheres.
Very recently Voshchinnikov et al. (2006) have studied the effect
of grain porosity on interstellar extinction, dust temperature,
infrared bands and millimeter opacity. They have used both, the
EMT-Mie based calculations and layered sphere model.

We have used discrete dipole approximation (DDA) to study
the extinction properties of the composite grains. For the
description on the DDA see Draine (1988).
The DDA allows the consideration of irregular shape effects,
surface roughness and internal structure within the grain
(Wolff et al. 1994, 1998 and Voshchinnikov et al. 2005).
For discussion and comparison of DDA and EMT methods, including
the limits of the effective medium theory, see
Bazell and Dwek (1990), Perrin and Lamy (1990), Perrin and Sivan
(1990), Ossenkopf (1991) and Wolff et al (1994).
In our earlier study we had used composite spherical grain models
to  evaluate the interstellar extinction curve in the wavelength range
0.55--0.20$\mu m$ (Vaidya et.al. 2001).

In the present study, we use more realistic composite spheroidal grain
models and calculate the extinction efficiencies in the extended
wavelength region, 3.40--0.10$\mu m$ and linear polarization in
the visible - near infrared region, i.e. 0.35--1.00$\mu m$.

Using these extinction efficiencies of the composite grains with a
power law type grain
size distribution we evaluate the interstellar extinction
curve and linear polarization.
In addition to reproducing the observed interstellar extinction curve,
the grain model should also be consistent with the abundance constraints.
We estimate the volume extinction factor, an important parameter
from the point of view of the cosmic abundance,  for the composite
grain models that reproduce the average observed extinction.
 In section 2 we give the
 validity criteria for the DDA and the composite grain models.
In section 3 we present the results of our computations and discuss
them.
The main conclusions of our study are given in section 4.

\section[]{Discrete Dipole Approximation (DDA) and Composite grains}

The basic DDA method consists of replacing a particle by an array of N oscillating
polarizable point dipoles (Draine, 1988). The dipoles are located on a lattice
and the polarizability is related to the complex refractive index $m$ through a
lattice dispersion relationship (Draine \& Goodman, 1993). Each dipole responds
to the external electric field as well as to the electric field of the other
N-1 dipoles that comprise the grain. The polarization at each dipole site is 
therefore coupled to all other dipoles in the grain.

In the present study, we have used the ddscat6.1 code (Draine \& Flatau, 2003) which
has been modified and developed by Dobbie (1999) to generate the composite grain
models. The code, first carves out an outer sphere (or spheroid) from a lattice
of dipole sites. Sites outside the sphere are vacuum and sites inside are
assigned to the host material. Once the host grain is formed, the code locates
centers for internal spheres to form inclusions. The inclusions are of a single
radius and their centers are chosen randomly. The code then outputs a three
dimensional matrix specifying the material type at each dipole site which is then
received by the ddscat program. In the present case, the sites are either
silicates, graphite or vacuum.

Using the modified code, we have studied composite grain models with a host silicate
spheroid containing number of dipoles 
N=9640, 25896 and 14440, each carved out from
$32 \times 24 \times 24$, $48 \times 32 \times 32$
and  $48 \times 24 \times 24$ dipole sites, respectively;
sites outside the spheroid are set to be vacuum and sites inside are
assigned to be the host material.
 It is to be noted that the composite spheroidal grain with
N=9640 has an axial ratio of 1.33, whereas N=25896 has the axial
ratio 1.5, and N=14440 has the axial ratio 2.0.
The volume fractions of the graphite inclusions used are
10\%, 20\% and 30\% (denoted as f=0.1, 0.2 and 0.3)
Details on the computer code and the corresponding modification
to the ddscat code (Draine \& Flatau 2003) are given in Dobbie
(1999), Vaidya et al. (2001) and Gupta et al. (2006).
Figure 1 and 2 illustrate the composite grain model with number of dipoles
N=9640 for the host spheroid and eleven inclusions. 

\begin{figure}
\includegraphics[width=84mm]{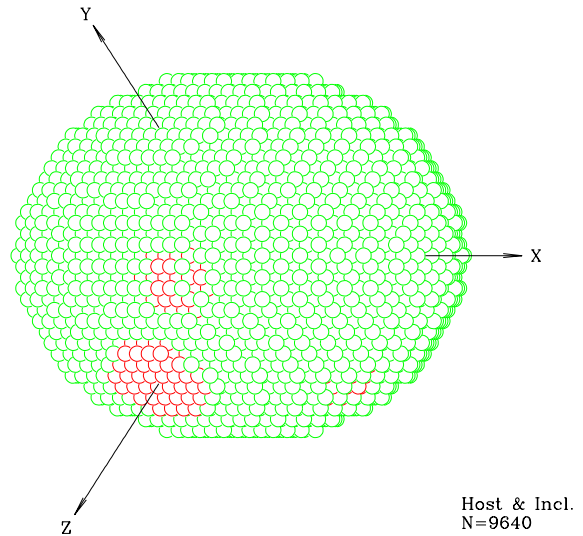}
\caption{A typical non-spherical composite grain with a
total of N=9640 dipoles where the inclusions
embedded in the host spheroid are shown such that only the ones placed
at the outer periphery are seen.}
\end{figure}

\begin{figure}
\includegraphics[width=84mm]{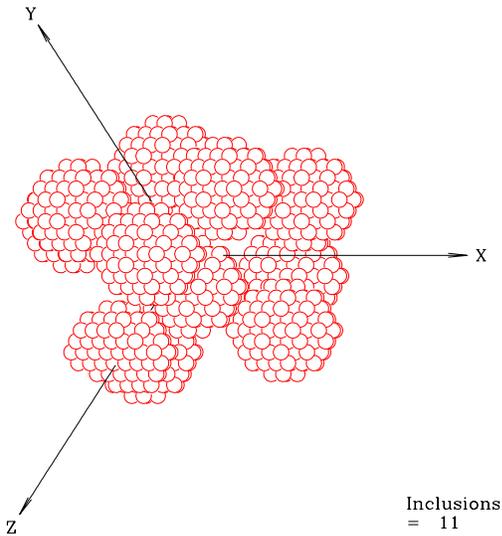}
\caption{Same as Fig. 1 but shows the inclusions.}
\end{figure}

Table 1 shows the number
of dipoles for each grain model (first column), 
number of dipoles per inclusion with the number of inclusions
denoted in bracket for volume fraction f=0.1 (second column). The
third and fourth column are the corresponding values for the remaining
volume fractions i.e. f=0.2 and 0.3.

\begin{table*}
\begin{center}
\caption{Number of dipoles for each inclusion for the three model}
\begin{tabular}{lccc}
\hline
No. of Dipoles(Axial ratio) & f=0.1 & f=0.2 & f=0.3\\
\hline
N=9640(1.33) & 152(6) & 152 (11) & 152(16)\\
N=25896(1.50) & 224(6) & 224 (11) & 224(16)\\
N=14440(2.00) & 432(7) & 432 (13) & 432(19)\\
\hline
\end{tabular}
\end{center}
\end{table*}

There are two validity criteria for DDA (see e.g. Wolff et al. 1994);
viz. (i) $\rm |m|kd \leq 1$, where m is the complex refractive index
of the material, k=$\rm \pi/\lambda$ is the wavenumber and
 d is the lattice dispersion spacing and
(ii) d should be small enough (N should be sufficiently large) to
  describe the shape of the particle satisfactorily.
 The complex refractive indices
 for silicates and graphite are obtained from Draine (1985, 1987).
For any grain model, the number of dipoles required to
obtain a reliable computational result can be estimated using the
ddscat code (see Vaidya \& Gupta 1997 and 1999, Vaidya et al. 2001).
For the composite grain model, if the host grain has N dipoles,
its volume is N(d)$^3$ and if 'a' is the
 radius of the host grain , N(d)$^3$=4/3$\rm \pi(a)^3$,
hence, N=4$\rm \pi/3(a/d)^3$, and if $\rm |m|kd \leq 1$ and
k=$\rm \pi/\lambda$
the number of dipoles N can be estimated at a given wavelength
and the radius of the host grain.
For all the composite grain models, with N=9640, 25896 and
14440 and for all the grain sizes, between a=0.001--0.250$\mu$, in
the wavelength range of 3.40--0.10$\mu m$, considered in the present
study; we have checked that
the DDA criteria are satisfied.

Table 2 shows the maximum grain size 'a' that satisfies the DDA validity
criteria at several wavelengths for the composite grain models with
N=9640, 14440 and 25896.

\begin{table*}
\begin{center}
\caption{DDA validity criteria}
\begin{tabular}{lccc}
\hline
$\lambda$ ($\mu m$)& N=9640 & 14440 & 25896\\
 & a($\mu$) & a($\mu$) & a($\mu$)\\
\hline
 
3.4000 & 4.00 & 5.00 & 6.00\\
2.2000 & 2.50 & 3.50 & 4.00\\
1.0000 & 1.20 & 1.40 & 1.60\\
0.7000 & 0.80 & 1.20 & 1.00\\
0.5500 & 0.60 & 0.96 & 0.80\\
0.3000 & 0.40 & 0.50 & 0.45\\
0.2000 & 0.22 & 0.30 & 0.25\\
0.1500 & 0.14 & 0.20 & 0.16\\
0.1000 & 0.10 & 0.16 & 0.12\\
\hline
\end{tabular}
\end{center}
\end{table*}

It must be noted here that the composite spheroidal grain
models with N=9640, 25896 and 14440 have the axial ratio 1.33,
1.5 and 2.0 respectively and if the semi-major axis and 
semi-minor axis are denoted by x/2 and y/2 respectively, then
$\rm a^3=(x/2)(y/2)^2$, where where 'a' is the
radius of the sphere whose volume is the same as that of
a spheroid.
In order to study randomly oriented spheroidal grains, it is
necessary to get the scattering properties of the composite
grains averaged over all of the possible orientations; in the
present study we use three values for each of the orientation
parameters ($\rm \beta, \theta and \phi$), i.e. averaging over 27 orientations,
which we find quite adequate (see e.g. Wolff et al. 1998).

\section[]{Results}

\subsection{Extinction Efficiency of Composite Spheroidal Grains}

Earlier, we had studied the extinction properties of composite
grains made up of the host spherical silicate grains with graphite
inclusions in the limited wavelength region 0.55--0.20$\mu m$
(Vaidya et al. 2001).
However, since the observed interstellar polarization requires that
the interstellar grains must be nonspherical, in the present paper we
study the extinction properties and linear polarization of the
 composite spheroidal grains
with three axial ratios, viz. 1.33, 1.5 and 2.0, corresponding to
the grain models with number of dipoles N=9640, 25896 and 14440 respectively,
 for three volume fractions of inclusions; viz. 10\%, 20\% and 30\%,
in the extended wavelength region 3.40--0.10$\mu m$.

Figures 3 (a-f) show the extinction efficiencies ($\rm Q_{ext}$) for the
 composite grains with the host silicate spheroids containing 9640,
 25896 and 14440 dipoles, corresponding to axial ratio 1.33, 1.5
 and 2.0 respectively.
The three volume fractions, viz. 10\%, 20\% and 30\%, of graphite
inclusions are also listed in the top (a) panel.
The radius of the host composite grain is set to 0.01$\mu$ for
all the cases.
The extinction in the spectral region 0.55--0.20$\mu m$ is highlighted
in the panels (d), (e) and (f).

\begin{figure}
\includegraphics[width=84mm]{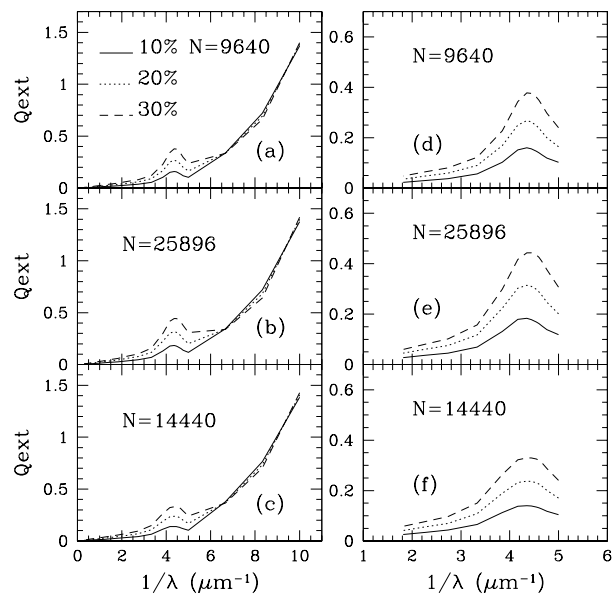}
\caption{Extinction Efficiencies for the composite grains of size
 0.01$\mu$
 with host spheroids containing dipoles N=9640, 25896 and 14440 are
 shown in (a),(b) and (c) in the wavelength region, 3.40--0.10$\mu m$.
 The panels (d),(e) and (f) show the extinction curves in the wavelength
region 0.55--0.20$\mu m$.}
\end{figure}

The effect of the variation of volume fraction of inclusions is
clearly seen for all the models. The extinction efficiency increases as
the volume fraction of the inclusion increases.
It is to be noted that the wavelength of the peak extinction
shifts with the variation in the volume fraction of inclusions.
These extinction curves also show the variation in the width
of the extinction feature with the volume fraction of inclusions.
All these results indicate that the inhomogeneities within the
grains play an important role in modifying the '2175\AA~' feature.
 Voshchinnikov (1990) and Gupta et al.
(2005) had found
variation in the '2175\AA~' feature with the shape of the grain, and
Iati et al. (2001, 2004); Voshchinnikov (2002);
Voshchinnikov and Farafonov (1993) and Vaidya et al. (1997, 1999)
had found the variation in the feature with the porosity of the grains.
Draine \& Malhotra (1993) have found relatively little effect on either
the central wavelength or the width of the feature for the coagulated
graphite silicate grains.

We have also computed the extinction efficiencies
of the composite spheroidal grains using the EMT-T-matrix based calculations. 
These results are
displayed in Figures 4 (a-c). For these calculations, the optical constants
were obtained using the Maxwell-Garnet mixing rule (i.e.
effective medium theory, see Bohren and Huffman 1983). Description
of T-matrix method/code is given by Mishchenko (2002).  
The extinction curves obtained using the
EMT-T-matrix calculations, deviate from the extinction curves obtained using 
the DDA, particularly in the 'bump region', 
i.e. 0.55--0.20$\mu m$. In Figures 5 (a-c) we have plotted the 
ratio Q(EMT)/Q(DDA) to compare the results obtained by both methods.
The results based on the EMT-T-matrix calculations and DDA results do not agree 
because the EMT does not take into account the inhomogeneities within the
grain; (viz. internal structure, surface, voids) (see Wolff et al. 1994, 1998)
and material interfaces and shapes are smeared out into a homogeneous 'average
mixture' (Saija et al. 2001). However, it would still be very useful and desirable
to compare the DDA results for the composite grains with those computed by other
EMT/Mie type/T matrix techniques in order to examine the applicability
of several mixing rules. (see Wolff et al. 1998, Voshchinnikov and Mathis
1999, Chylek et al. 2000, Voshchinnikov et al. 2005, 2006). The application of DDA, 
poses a computational challenge, particularly for the large values of
the size parameter X ($\rm=2\pi a/\lambda > 20$ ) and the 
complex refractive index m 
of the grain material would require large number of dipoles and that in
turn would require considerable computer memory and cpu time (see e.g. Saija et al.
2001, Voshchinnikov et al. 2006). 
 
\begin{figure}
\includegraphics[width=84mm]{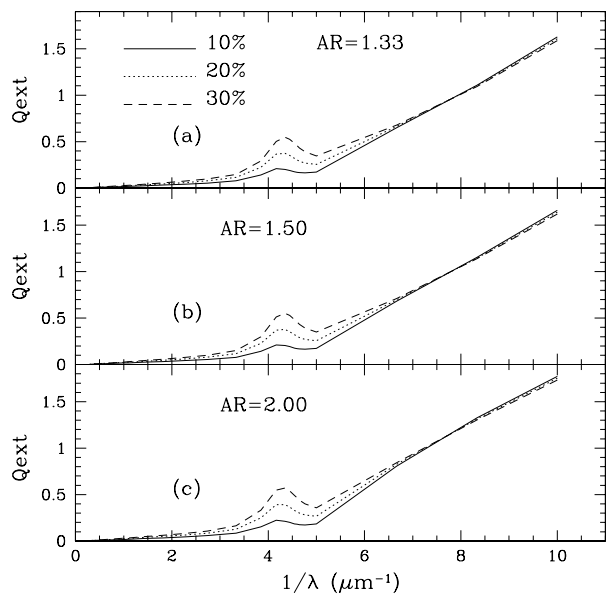}
\caption{Extinction Efficiencies for the composite spheroidal grains of size
 0.01$\mu$ with three axial ratios (AR=1.33, 1.5 and 2.0)
using EMT-T Matrix based calculations in the wavelength region
3.4--0.10$\mu m$.}
\end{figure}

\begin{figure}
\includegraphics[width=84mm]{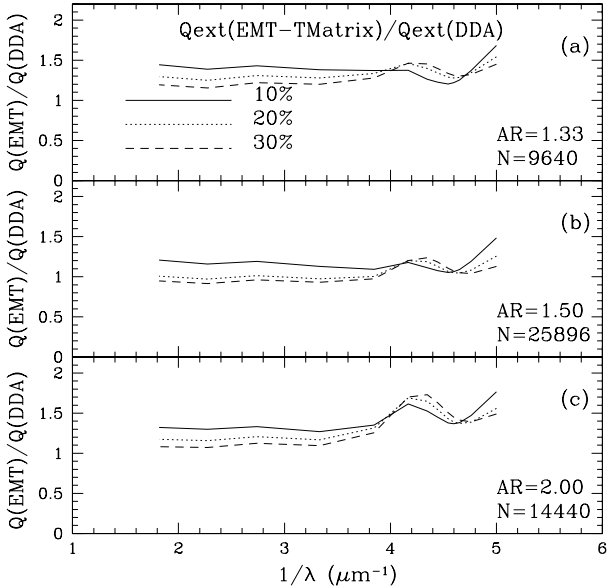}
\caption{
Ratio Q(EMT-Tmatrix)/Q(DDA) in the wavelength region 0.55-0.20$\mu m$
for the composite spheroidal grains of size 0.01$\mu$ with three axial ratios,
AR=1.33, 1.5, 2.0 corresponding to N=9640, 25896 and 14440 respectively.} 
\end{figure}

Mathis \& Whiffen (1989), Mathis (1996) and Voshchinnikov et al. (2006)
 in their composite grain
models have used amorphous carbon with silicate. We have not considered
it in the present study as amorphous carbon particles exhibit absorption
at approximately 2500\AA~ and also it is highly absorbing at very long
wavelengths and would provide most of the extinction longward of
0.3$\mu m$ (Draine 1989, Weingartner and Draine 2001). It
is also not favoured by Zubko et al. (2004). Instead, large PAH
molecules are likely candidates to be the carrier of the interstellar
2175\AA~ feature -- a natural extension of graphite hypothesis
(Draine, 2003b). 

Figures 6(a-d) show the extinction efficiencies ($\rm Q_{ext}$) for the
composite grains for four host grain sizes: viz. a=0.01, 0.05,
0.1 and 0.2 $\mu$ at a constant volume fraction of inclusion of 20\%.

 It is seen that the extinction and the shape of the extinction curves
varies considerably as the grain size increases. The '2175\AA~ feature'
is clearly seen for small grains ; viz. a=0.01 and
0.05$\mu$, whereas for larger grains the feature almost disappears.
It is also to be noted that there is no appreciable variation in the
extinction with the axial ratio of spheroidal grains; i.e. 
1.33, 1.5, 2.0 corresponding to N=9640, 25896 and 14440.
 
\begin{figure}
\includegraphics[width=84mm]{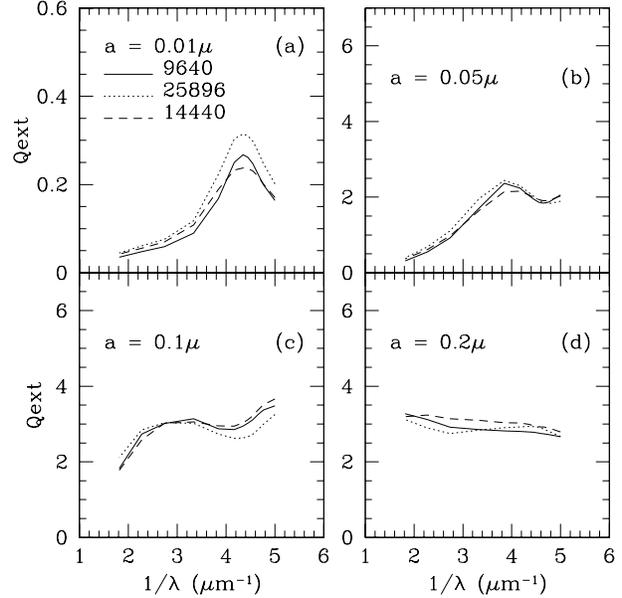}
\caption{Extinction efficiencies for the composite grains with
various sizes with 20\% volume fraction of graphite inclusions.}
\end{figure}

\subsection{Interstellar Extinction Curve}

The interstellar extinction curve (i.e. the variation of extinction
with wavelength) is usually expressed by the ratio
$\rm E(\lambda - V)/E(B - V)$ versus $1/\lambda$.
We use the extinction efficiencies of the composite grains,
with a power law size distribution (i.e. $\rm n(a) \sim a^{-3.5}$,
 Mathis et al. 1977) to evaluate the interstellar
extinction curve in the wavelength region of 3.40--0.10$\mu m$.
In addition to the composite grains a separate component of small
graphite grains is required to produce the observed
peak at 2175\AA~ in the interstellar extinction curve (Mathis,
1996). It must also be mentioned here that the most widely accepted
explanation of the 2175\AA~ bump has been the extinction by small 
($\sim a < 0.05 \mu$)
graphite grains (e.g. Hoyle and Wickramasinghe 1962, Mathis et al. 1977, Draine 1989).
Also, the stability of the observed feature at 2175\AA~ along all the lines of sight
rules out the possibility of using composite grains, made up of silicate
with graphite as inclusions to reproduce the feature.
(Iati et al. 2001).

The average observed interstellar extinction curve 
(Savage and Mathis 1979; Whittet, 2003)
is compared with with the model curve formed from a $\chi^2$ minimized
and best fit linear combination of the composite grains (contributory fraction x)
 and graphite grains (contributory fraction y); i.e the model interstellar 
extinction curves for the composite
 grains and the graphite grains are linearly combined to render a net curve
 for comparison with the
average observed extinction curve.
 The formula to obtain the minimized $\chi^2$ values
is given by Bevington (1969).

$$   
{\chi{^2_j}} = \frac {\sum_{i=1}^n (S_{i}^j-T_{i}^k)^2} {pp}  \eqno (1)
$$

where pp is the degrees of freedom, $S_{i}^{j}(\lambda_{i})$ 
is the $j$th model curve for the 
corresponding $x$ and $y$ linear combination of composite and graphite 
grains and $T_{i}^{k}(\lambda_{i})$ is for the observed curve, 
$\lambda_{i}$ are the wavelength points with i=1,n 
where n are the number of wavelength points of the extinction curves.
Details are given in our earlier papers (see Vaidya \& Gupta 1999, 
Vaidya et al. 2001).

Table 3 shows the best fit $\chi^2$ values for the extinction curves
for the composite grain models with volume fraction of
inclusions f=0.1, 0.2, 0.3 for three wavelength ranges, viz. 3.40--0.10$\mu m$,
3.40--0.55$\mu m$ and 0.55--0.20$\mu m$. The numbers in the brackets (x/y)
adjacent to 
each $\chi^2$ value is the fractional contibution of the composite Si+f*Gr and
the required additional small graphite grain e.g. (0.5/0.3) means that there
is 0.5 contribution from the composite grain and 0.3 contribution from this
additional graphite grain to obtain the corresponding minimum $\chi^2$ value.

\begin{table*}
\begin{center}
\caption{Best fit $\chi^2$ values for the Interstellar Extinction Curves
for the Composite Spheroidal grain models in the wavelength range 3.40--0.10$\mu m$,
3.40--0.55$\mu m$ and 0.55--0.20$\mu m$ with grain size distribution a=0.005--0.250$\mu$. The numbers in the brackets adjacent to
each $\chi^2$ value is the fractional contibution of the composite Si+f*Gr and
the required additional small graphite grain.}

\begin{tabular}{lccc} 
\hline 
Vol. fraction & N=9640 & N=25896 & N=14440\\
\hline
Wavelength range & 3.40--0.10$\mu m$ &  &  \\
\hline
 f=0.1 & 0.1635(0.5/0.3) & 0.1811(0.5/0.3) & 0.1659(0.5/0.3)\\
 f=0.2 & 0.2045(0.5/0.3) & 0.2483(0.5/0.3) & 0.1839(0.5/0.3)\\
 f=0.3 & 0.3053(0.5/0.3) & 0.4532(0.5/0.3) & 0.3115(0.5/0.3)\\
\hline
Wavelength range & 3.40--0.55$\mu m$ &  &  \\
\hline
 f=0.1 & 0.0148(0.5/0.3) & 0.0148(0.6/0.2) & 0.0176(0.5/0.3)\\
 f=0.2 & 0.0273(0.7/0.1) & 0.0352(0.6/0.1) & 0.0306(0.7/0.1)\\
 f=0.3 & 0.0360(0.6/0.1) & 0.0570(0.6/0.1) & 0.0400(0.6/0.1)\\
\hline
Wavelength range & 0.55--0.20$\mu m$ &  &  \\
\hline
 f=0.1 & 0.0672(0.4/0.4) & 0.0899(0.4/0.4) & 0.0766(0.6/0.3)\\
 f=0.2 & 0.1192(0.3/0.4) & 0.1578(0.3/0.4) & 0.1028(0.4/0.4)\\
 f=0.3 & 0.1376(0.3/0.4) & 0.1658(0.3/0.4) & 0.1364(0.4/0.4)\\
\hline
\end{tabular}
\end{center}
\end{table*}

Figure 7 shows the interstellar extinction
curves for the composite grain models with number of dipoles for
the host spheroids N=9640, 25896 and 14440 and volume fractions of inclusions
f=0.1, 0.2 and 0.3 in the entire wavelength region of 3.40--0.10$\mu m$ for
the power law grain size distribution, $\rm n(a) \sim a^{-3.5}$,
in the size range, a=0.005--0.250$\mu$.

\begin{figure}
\includegraphics[width=84mm]{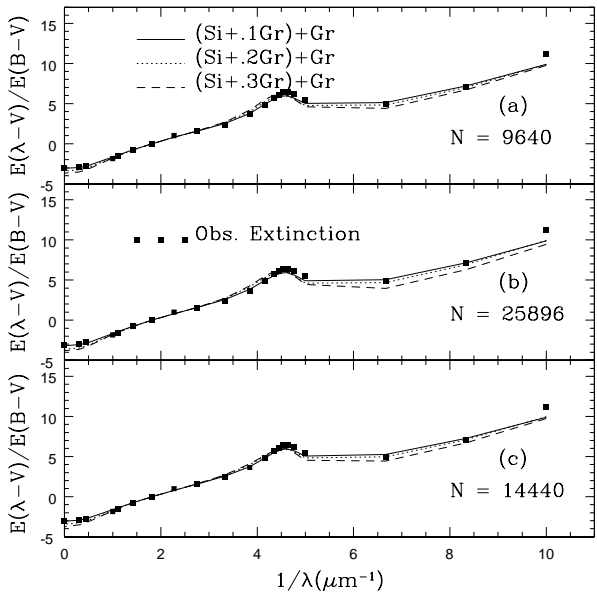}
\caption{Comparison of the observed interstellar extinction curve
with the best fit model combination curve of composite grains with
three volume fractions of graphite inclusions (N=9640, 25896 and
14440)
and graphite grains in the wavelength range of 3.40--0.10$\mu m$.}
\end{figure}

It is seen from Figure 7 and Table 3 that the composite spheroidal
grain models with N=9640 and f=0.1 fit the average observed extinction
curve quite satisfactorily in the entire wavelength range considered,
i.e 3.40--0.10$\mu m$, in this study
The model extinction curves with N=25896, 14440 deviate from the observed
extinction curve in the uv region, i.e. beyond the wavelength
$\sim$ 0.1500$\mu m$ (i.e. 6$\mu m^{-1}$).
These results indicate that in addition to composite grains and graphite,
a third component of very small grains (e.g very small silicate grains or PAHs) 
may be required to explain the extinction beyond 1500\AA~ in the UV 
(Weingartner and Draine, 2001).

Figure 8 shows the extinction curves in the wavelength range
0.55--0.20$\mu m$ for the composite grain models.
It is seen that all the model curves fit quite well with
the observed interstellar extinction curve in this wavelength region.
$\chi^2$ values are also quite low in this region (see Table 3).

\begin{figure}
\includegraphics[width=84mm]{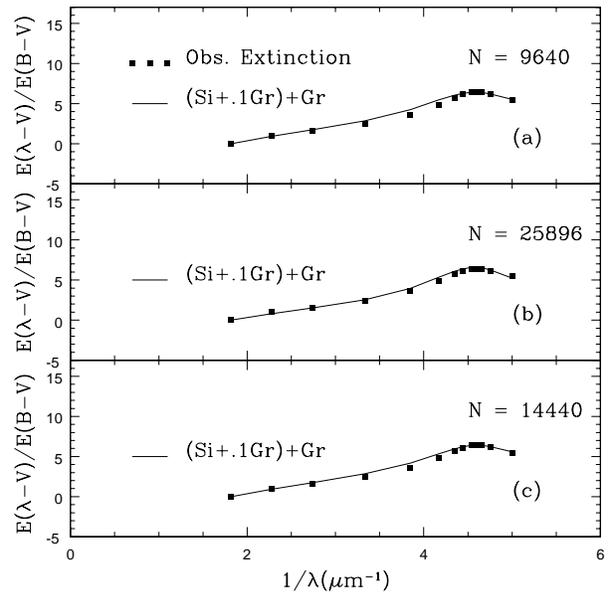}
\caption{Comparison of the observed interstellar extinction curve
with the best fit model combination curve of composite grains
(N=9640, 25896 and 14440) and graphite grains in the wavelength
range of 0.55--0.20$\mu m$.}
\end{figure}

We have also evaluated extinction curves for the smaller size
range, viz. a=0.001--0.100$\mu$, so that the DDA validity criteria
is satisfied for the grain models with N=9640 in the uv spectral
region (see Table 2).
Figure 9 shows the interstellar extinction curves for the composite
grain models with N=9640 in the size range a=0.001--0.100$\mu$.  
The $\chi^2$ values for these model curves are 0.0908, 0.1094 and
0.1425 for the volume fractions f=0.1, 0.2 and 0.3 respectively.

\begin{figure}
\includegraphics[width=84mm]{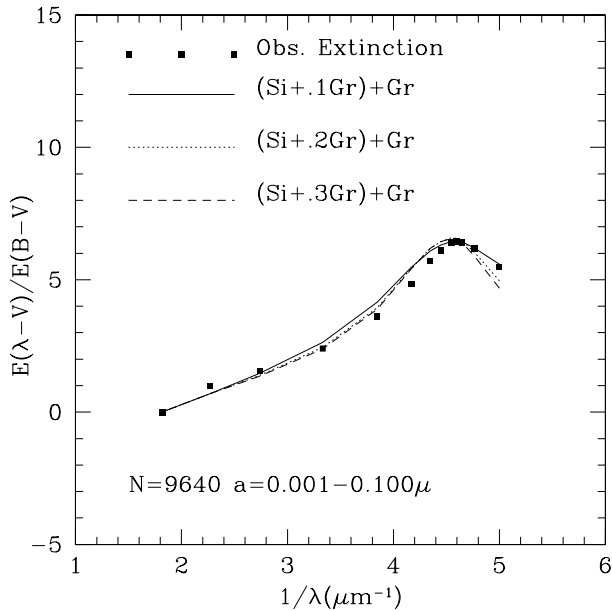}
\caption{Extinction curves for composite grain models with N=9640
for the size range, a=0.001--0.100$\mu$}
\end{figure}

These results show that the composite spheroidal grain models with the axial
ratio of the host silicate spheroid not very large; i.e $\sim 1.33$,
N=9640
and the volume fraction of the graphite inclusions, f=0.1 fit the observed
extinction satisfactorily
in the entire wavelength range 3.40--0.10$\mu m$, whereas in the wavelength range
0.55--0.20$\mu m$, all the composite spheroidal grain models with
N=9640, 25896 and 14440 fit the observed extinction
curve better and the $\chi^2$ values are lower.

Zubko et al. (1996, 1998) have used multicomponent mixtures of bare spherical
grains to analyze the interstellar extinction curves. They have used the
method of regularization for this analysis.
Recently Iati et al. (2004), Zubko et al. (2004)
Voshchinnikov et al. (2005) and Maron \& Maron (2005) 
have also proposed composite grain models. However, all these
authors have used EMT to obtain the optical constants for the
composite grain models. Andersen et al. (2002) have performed
extinction calculations for clusters of polycrystalline graphite
and silicate spheres, using discrete dipole approximation.
Very recently 
Voshchinnikov et al. (2006) have used both EMT-Mie type and layered sphere based
 calculations for the composite porous grain models. Voshchinnikov et al. (2006)
have found the model extinction curves obtained using layered sphere based
calculations fit the observed extinction better.   

\subsection {Linear Polarization}

The linear polarization curve, usually
plotted as $\rm P_{\lambda}$ versus $1/\lambda$, displays a broad peak in the
visible region for most stars and the wavelength of maximum polarization
$\rm \lambda_{max}$, varies from star to star, with a mean value
at around 0.55$\mu m$.
The dependence of the linear polarization on the wavelength is described
by the empirical formula (Serkowski et al. 1975, Whittet 2003);

$\rm P_{\lambda}/P_{max}=exp [-Kln^2(\lambda/\lambda_{max})]$

where $\rm P_{max}$ is the degree of polarization at the peak, and
the parameter K, determines the width of the peak.
This formula with K=1.15 provides an adequate representation of the
observations of interstellar polarization in the visible-NIR region
(0.36--1.00$\mu m$) (Whittet et al. 1992).
It is also important to note that
the wavelength dependence of interstellar
polarization is a function not only of the size, shape and
composition of the
dust grain but also of orientation of the grains (see e.g. Wolff et al.
1993).

Using ddscat (Draine and Flatau 2003)
we have calculated linear polarization efficiency,
$\rm |Q_{pol}|=Q_{ext}(E) - Q_{ext}(H)$
for the aligned composite spheroidal
grains at several orientation angles;
where $\rm Q_{ext}(E)$ and $\rm Q_{ext}(H)$
are extinction efficiency factors for the directions of the incident
field vector Q(E) and perpendicular Q(H) to the axis of the spheroid.
In this paper we have restricted the polarization study to
the wavelength region between 1.00--0.30$\mu m$.

In Figure 10 we show the extinction efficiency $\rm Q_{ext}(E)$ and $\rm Q_{ext}(H)$
for the composite grain models N=9640, f=0.1 at three orientation angles.

We carried out the linear polarization calculations with MRN-type power law
grain size distribution by varying the power law index from p = -1.5 to -4.0
and the results are shown in the Figure 11 along with the Serkowiski's curve.
It may be noted that the power law index p=-2.3 and -2.5 fit the Serkowski's 
curve reasonably well.
Figures 12(a) and (b) show the linear polarization for the composite grain
models with N=9640; f=0.1 and 0.05 respectively for
a MRN-type grain size distribution with power law index p=-2.5,
compared with the curve derived from Serkowski's formula (Whittet 2003).
It is seen that composite spheroidal
grain models with smaller fraction of graphite inclusions, i.e. f=0.05
 fit better with Serkowski's curve.
It is also seen that the results with $\theta=90^{\circ}$ fit the
Serkowski's curve the best. Our results are consistent with that
pointed out by Mathis (1979) and Wolff
et al. (1993), i.e. for the interstellar polarization curve, the model fit parameters
including the size distribution, are quite different from those parameters
required to fit the extinction curve. Mathis (1979) required a power
law index p=-2.5 and Wolff et al. (1993) required p=-2.7 to fit the Serkowski's
curve. Wolff et al. (1993) have further noted that the MRN model requires altering
the size distribution to fit the Serkowski's curve.

These results on the composite spheroidal grains with silicate and
graphite as constituent materials also indicate that most of the polarization is
produced by the silicate material. Our results are in agreement
with the results obtained by Mathis (1979) and Wolff et al. (1993).
Duley et. al. (1989) have used a core-mantle grain model consisting of
silicate as core and hydrogeneted amorphous carbon (HAC) as mantle and have
shown that polarization is mostly produced by silicate.

It must be noted here that the two most important parameters
characterizing the extinction and polarization curves are: viz.
(i) the ratio $\rm R [=A_{v}/E(B-V)]$ of total to selective extinction and
(ii) $\rm \lambda_{max}$;
 and a linear correlation exists between R and $\rm \lambda_{max}$,
 given by R=5.6$\rm \lambda_{max}$, (Whittet 2003).

The observed ratio of polarization, $\rm P_{V}$, to the extinction,
$\rm A_{V}$ i.e. $\rm P_{V}/A_{V}$ is generally 0.025 but
higher value, viz. 0.06, is also observed (Greenberg 1978).

We have calculated  $\rm P_{V}/A_{V}$ and $\rm \lambda_{max}$  for
the composite spheroidal grain models that fit the Serkowski's curve viz.
Figure 11. These results are shown in Table 4. It is seen that the grain models
with N=9640 and f=0.05 are consistent with the observed values 
i.e. $\rm \lambda_{max} = 0.55 \mu m$ and $\rm P_{V}/A_{V} = 0.02$.
In the present study, we have not discussed the mechanism for the alignment
of the grains.

\begin{table*}
\begin{center}
\caption{Interstellar Linear Polarization parameters for composite spheroidal
grain models.}
\begin{tabular}{lccc} \hline 
Si+Gr Models & $\theta$ & $P_{V}/A_{V}$ & $\lambda_{max} (\mu m)$\\
\hline
f=0.1 & & & \\
\hline
N=9640 & $45^{\circ}$ & 0.007 & 0.44 \\
N=9640 & $60^{\circ}$ & 0.011 & 0.44 \\
N=9640 & $90^{\circ}$ & 0.018 & 0.55 \\
\hline
f=0.05 & & & \\
\hline
N=9640 & $45^{\circ}$ & 0.012 & 0.37 \\
N=9640 & $60^{\circ}$ & 0.019 & 0.55 \\
N=9640 & $90^{\circ}$ & 0.025 & 0.55 \\
\hline
\end{tabular}
\end{center}
\end{table*}

\begin{figure}
\includegraphics[width=84mm]{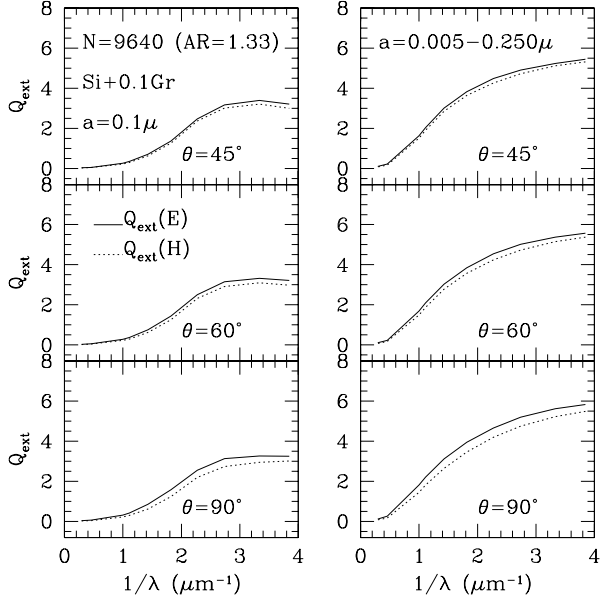}
\caption{Extinction Efficiency for composite grain model with N=9640
and f=0.1 at three orientation angles. The curves on the left panel are for 
a single size grain a=0.1$\mu$ and the ones on the right are for
size distribution range a=0.005-0.250$\mu$.}
\end{figure}

\begin{figure}
\includegraphics[width=84mm]{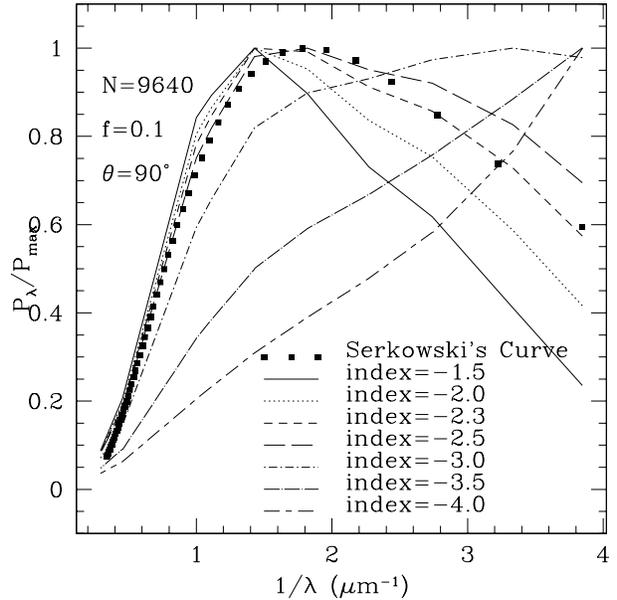}
\caption{Linear Polarization curves for composite grain models and fitting
with the Serkowski's curve with various power law indices.}
\end{figure}

\begin{figure}
\includegraphics[width=84mm]{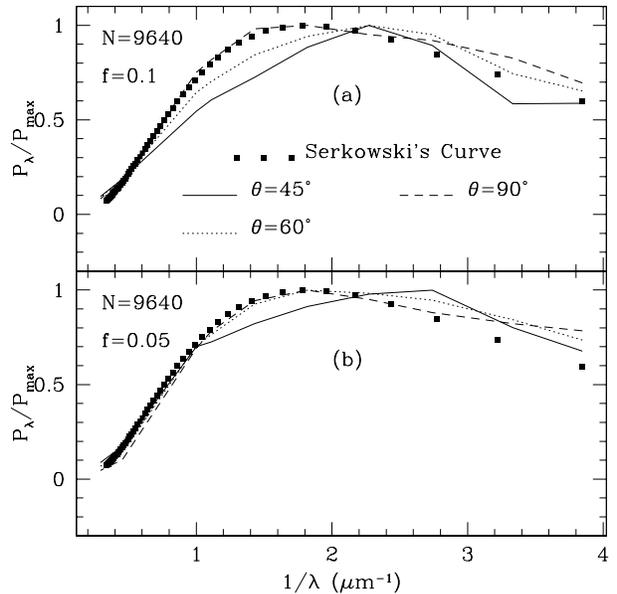}
\caption{Linear Polarization for Composite Spheroidal Grains for N=9640 and
25896 with volume fraction of graphite f=0.1 and f=0.05, compared with
Serkowski's Law.}
\end{figure}

 \subsection{Volume Extinction Factors and Cosmic Abundances}

In addition to reproducing the interstellar extinction curve any
grain model must also be consistent with the abundance constraints.
Snow and Witt (1995, 1996) have reviewed several models for the
interstellar dust,which provide the data on the quantities of some
elements that are required to reproduce the interstellar extinction.
They have found that there is not only a carbon crisis
(Kim \& Martin, 1996) but there are now tight constraints on other elements
as well and almost all models require about 1.5--2.0 times more
silicon than that is available.
Mathis (1996) and Dwek (1997) have proposed composite fluffy dust
models (CFD) to overcome the cosmic abundance constraints.
Using the composite grains of silicates and amorphous carbon
Mathis (1996) has obtained the cosmic carbon abundance of C atoms (per
$10^{6}$ atoms), C/H, of about 140--160.
However, Mathis has used EMT to obtain
optical constants for the composite grains and then used Mie theory to
calculate extinction cross sections, which were then multiplied by a factor
1.09 to account for the enhancements in the extinction for the
nonspherical grains. Recently, Zubko et al.. (2004) have also used
EMT/Mie theory to study the optical properties of composite grains.
This approach is found to be questionable (Saija et al. 2001, Weingartner
and Draine, 2001). In our earlier study on the composite spherical grains
(Vaidya et al. 2001) as well as in the present study on the composite spheroidal
grains (see Figure 5) we have shown the inherent inability of
EMT based calculations to treat the scattering/extinction by composite
grains. Wolff et al. (1993) have also noted that the composite grain
model using EMT cannot achieve a meaningful fit to the observed data.
Also, the use of the  'Be' amorphous carbon in the
composite model is not favoured as it is much more absorbing
at long wavelengths and would provide most of the extinction
for all wavelengths $>0.3\mu m$ (Weingartner and Draine, 2001).
We have used the more accurate DDA method to calculate the
extinction cross sections for the composite grains, made up of the host
silicate spheroids and inclusions of graphite and have showed that
 the composite grain models are more
efficient than the bare grains, containing single component,
 in producing interstellar extinction.

 An important parameter from the point of view of cosmic abundance
is the volume extinction factor $\rm V_{c}$, defined as the ratio of the total
volume of the grains to the total extinction cross-section of the
grains i.e. $\rm \sum V/\sum C_{ext}(\lambda)$ (Greenberg \& Hong 1975;
Vaidya et al. 1984 and Gupta et al. 2005).
It is to be noted here that $\rm V_{c}$  directly determines the amount of
material required to
 produce the extinction at a specific wavelength.
Table 3 shows the volume extinction factor $\rm V_{c}$ for the
composite spheroidal grain models at $\rm \lambda=0.55\mu m$.

 \begin{table*}
 \begin{center}
\caption{Volume Extinction factors V$_{c}$ for Composite Spheroidal Grains
and Cosmic Abundances.}
\begin{tabular}{lccc} 
\hline
 Si+Gr Models & f=0.1 & f=0.2 & f=0.3\\
\hline
N=9640 & 0.209 & 0.180 & 0.159 \\
N=25896 & 0.199 & 0.165 & 0.145 \\
N=14440 & 0.207 & 0.175 & 0.152 \\
\hline
N=9640 & C/H,Si/H & C/H,Si/H & C/H,Si/H\\
(ppm)  & 160,28   & 170,26  & 180,24\\
\hline
\end{tabular}
\end{center}
\end{table*}

It is seen that for all the three volume fractions
of inclusions, viz.  f=0.1, 0.2 and f=0.3, the composite grain
model with N=25896 (axial ratio 1.5) is the most efficient in
producing the visual extinction. The volume extinction factor V$_{c}$ is
the lowest for this grain model.
It is important to note here, that these values of the volume extinction
factors for the
composite grain models, containing silicate as host and graphite as
inclusions,  are much lower than what we had
obtained for the bare silicate and graphite grain models (Gupta et al. 2005).
These results on the volume extinction factors clearly indicate
 that the composite grains are more efficient in producing the extinction
 i.e. the amount of silicate and graphite required is less than
that would be required for the bare silicate/graphite models.
The number of atoms (in ppm) of the particular material tied up in grains
can be estimated if the atomic mass of the element in the grain material
and the density of the material are known
(see e.g. Cecchi-Pestellini et al. 1995 and Iati et al. 2001).
 From the composite grain models we have proposed,
we estimate C abundance i.e. C/H to be between 160-180 (including those
atoms that produce the 2175\AA~ feature), which is
considerably lower than what is predicted by bare silicate/graphite grain models 
(e.g. C/H=300 ppm, Mathis et al. 1977; C/H=254 ppm,
Li and Draine, 2001) but it is still
 significantly above the recent ISM value of 110 (Mathis 2000)
The estimated Si abundance from the composite grain model presented
here is between 24-28, which is higher than the ISM value of 17 ppm (Snow and
Witt 1996,
Voshchinnikov 2002) but it is lower than the other recent grain
models  ($\sim$ 32, Li and Draine, 2001).
 Recently Voshchinnikov et al. (2006)
have estimated very low values for C/H ($\sim$137) and Si/H ($\sim$8.8)
with their highly porous grain models.
In Table 5, we also show the estimated C/H and Si/H abundance values derived from
the composite grain model N=9640 which is the best fit model.
Snow (2000) has addressed the issues related
 to and the question of appropriate reference abundance standards
and has noted that no model for the dust extinction copes successfully with
the reduced quantities of available elements imposed by the revised cosmic
abundance standards and the consequent reductions in depletions.
Draine (2003a) has also pointed that
 the uncertainties in the gas-phase depletions and in the
 dust compositions are quite large and hence one should not
 worry about the dust models that contradict the abundance
 constraints, up to a factor of two. 
 Weingartner and Draine (2001)
 have used populations of separate silicate, graphite and Polycyclic
 Aromatic Hydrocarbons (PAHs)
 spherical grains to obtain extinction curves in the Milky Way,
Large Magellanic cloud
and Small Magellanic cloud. The composite grain models with
silicates,  graphite and a separate component of PAHs as
 constituent materials may further help
 to reduce the requirements to match the abundance constraints .
Recently, Piovan et. al (2006) have also noted that any realistic model of a dusty
ISM to be able to explain the UV-optical extinction and IR emission has to include
at least three components, i.e. graphite, silicate and PAHs.

\section{Summary and Conclusions}

Using the discrete dipole approximation (DDA)
we have studied the extinction properties of the composite
spheroidal grains, made up of the host silicate and graphite inclusions
in the wavelength region of 3.40--0.10$\mu m$.
We have also calculated the linear polarization in the 
wavelength region, 1.00--0.30$\mu m$. Our main conclusions from
this study are:

(1) The extinction curves for the composite spheroidal grains show the
    shift in the central wavelength of the extinction peak
 as well as variation in
    the width of the peak with the variation in the volume fraction of the
    graphite inclusions. These results clearly indicate that the
shape, structure and inhomogeneity in the grains play an important 
role in producing the extinction. 

We also note that the extinction efficiency in the 'bump region' for the composite
grains obtained with EMT deviate considerably from that obtained by DDA.

(2) The model extinction curves for the composite spheroidal grains with the axial
   ratio not very large ($\sim 1.33$, N=9640) and 10 \% volume
   fractions of graphite inclusions are found to fit the average observed
interstellar extinction
   satisfactorily. Extinction curves with other composite grain models
   with N=25896 and 14440 also fit the observed
   extinction curve reasonably well, however these model curves deviate from the
observed
   curves in the UV region, i.e. beyond about wavelength 1500\AA~. These
results
indicate that
   a third component of very small particles in the composite grains may
help improve
   the fit in the UV region (see e.g. Weingartner and Draine 2001).

(3) The linear polarization curves obtained for the composite grain models
with silicate as the host and very small volume fraction (f=0.05) 
of graphite inclusions
fit the Serkowski curve; which indicates that most of the polarization
is produced by the silicate material (see Duley et. al. 1989; Mathis \&
Whiffen 1989 and Wolf et. al. 1993). The ratio $\rm P_{V}/A_{V}$ for these composite
spheroidal grains is also consistent with the observed values.

(4) The volume extinction factor for the composite grain models with host silicate
and graphite inclusions, is lower than that is obtained for the bare 
silicate/graphite grain models (e.g. Mathis et al. 1977).
These results clearly show that composite grain model is
more efficient in producing the extinction and it would perhaps help
to reduce the cosmic abundance constraints.

Perets and Biham (2006) have recently noted that due to complexity of various
processes, viz. grain-grain collisions and coagulation; photolysis and alteration
by UV radiation, X-rays and cosmic rays etc., there is no complete model that
accounts for all the relevant properties of the interstellar dust grains.

We have used the composite spheroidal grain model to fit
the observed interstellar extinction and linear polarization.
 The IRAS and COBE observations have indicated
the importance of the IR emission as a constraint on interstellar dust
models
(Zubko et al. 2004). It would certainly strengthen the composite spheroidal
grain model further, if it can fit the IRAS observations as well as COBE
data on diffuse IR emission (Dwek 1997) .

\section*{Acknowledgments}

DBV and RG thank the organizing committee of the symposium,
Astrophysics of Dust, Estes Park,CO. USA for
 the financial support which enabled them to participate in the symposium.
Authors thank Profs. N. V.
Voshchinnikov and A.C. Andersen for their suggestions. We thank 
the reviewer for his constructive comments which has helped in improving the
quality of the paper. The DDSCAT
code support from B. T. Draine and P. J. Flatau is also acknowledged.
DBV thanks Center for Astrophysics and Space Astronomy (CASA),
Boulder CO. USA for inviting him and providing him all the facilities
and also IUCAA for its continued support.

\label{lastpage}
\end{document}